\documentstyle[12pt]{article}
\def\la{\mathrel{\mathpalette\fun <}}
\def\ga{\mathrel{\mathpalette\fun >}}
\def\fun#1#2{\lower3.6pt\vbox{\baselineskip0pt\lineskip.9pt
  \ialign{$\mathsurround=0pt#1\hfil##\hfil$\crcr#2\crcr\sim\crcr}}}
\relax
\begin{document}
\begin{tabbing}
\hskip 14.5 cm \= {SU-ITP-92-8}\\
\hskip 1 cm \>{\today}\\
\end{tabbing}
\thispagestyle{empty}
\begin{center}
{\Large\bf    STRINGS, TEXTURES, INFLATION}\\
\vskip 1cm
{\Large\bf  AND SPECTRUM BENDING}\\
\vskip 2 cm
{\bf Andrei Linde} \footnote{On leave from: Lebedev
Physical Institute, Moscow.\, \,
E-mail: LINDE@PHYSICS.STANFORD.EDU}\\
\vskip .3cm
 Department of Physics, Stanford University, Stanford, CA 94305\\
\vskip 3 cm
{\large ABSTRACT}
\end{center}
\begin{quote}

We discuss relationship between inflation and various models of
production of
density inhomogeneities due to  strings, global monopoles, textures
and other
topological and non-topological defects. Neither of these models
leads to a
consistent cosmological theory without the help of inflation.
However, each
of these models can be incorporated into inflationary cosmology. We
propose
 a model of inflationary phase transitions, which, in addition to
topological
 and non-topological defects, may provide adiabatic density
perturbations
 with a sharp maximum between the  galaxy scale $l_g$ and the horizon
 scale $l_H$.

\end{quote}
\vfill \newpage
\vfill\eject

Modern cosmology has two apparently contradictory goals. First of
all, one
should explain why our universe is so flat, homogeneous, isotropic,
why it does
not contain such defects as  monopoles, domain walls,  etc. Then one
should
explain why our universe is not {\it perfectly} flat, homogeneous,
isotropic,
and why the deviation from perfection is so small (${\delta\rho\over
\rho} \sim
{\delta T\over T} \sim 10^{-5}$).

In our opinion, it is  somewhat risky to suggest various  solutions
to the
second problem without having at least an idea of  how to solve the
first one.
Fortunately, a possible solution to the first problem is well known,
it is
inflation.  Now, ten years after the inflationary scenario was
suggested, we
still do not find any fundamental flaws in it. Nor have we found any
alternative solution to the first problem. (Actually, we are speaking
about ten
different problems, which are solved simultaneously by one simple
scenario; for
a review see \cite{MyBook}.) The only alternative solution of the
homogeneity
problem which I am aware of is based on quantum cosmology \cite{L91}:
It is
possible that the probability of quantum creation of the universe,
like the
probability of tunneling with bubble production,  is particularly
large for
spherically symmetric universes. However, even if it is true, we will
still
need something like inflation to make the newly born universe not
only
symmetric but also extremely large.

It came as a great bonus to inflationary cosmologists when it was
realized that
inflation may solve not only the first problem, but the second one as
well:
Quantum fluctuations produced during inflation lead to generation of
adiabatic
density perturbations with a flat (almost exactly scale independent)
spectrum
\cite{Pert}. Thus, inflation offers the most economical possibility
to solve
all cosmological problems by one simple mechanism.

However, Nature is not very economical in the number of different
cosmological
structures. Even though it may be possible to explain the origin of
all these
structures by one basic mechanism, one should keep in mind all other
possibilities, such as adiabatic and isothermal perturbations with a
non-flat
spectrum, strings \cite{ZelVil}, global monopoles
\cite{MyBook,GlobVil},
textures \cite{Text}, late time phase transitions \cite{Schramm},
etc.

Adiabatic perturbations with a flat spectrum is a natural consequence
of  many
inflationary models. Therefore, there is a tendency  to identify
perturbations
with a flat spectrum with  inflation\footnote{Actually, even in
simplest models
the spectrum of perturbations is  not absolutely flat. For example,
in the
theory $\lambda\varphi^4$, density perturbations grow by about 1/3 in
the
interval from the galaxy scale to the scale of horizon.}.  Some
authors do it
just for brevity, to distinguish between  inflation
without strings and textures and inflation with string and textures
\cite{Brand}.
However, some other authors, when advertising new types of density
perturbations,  represent them as a real  alternative to inflationary
cosmology,
see,  e.g., \cite{ST}. Even though such an
 attitude is understandable, we do not think that it is well
motivated. None
 of the new  mechanisms of generation of density perturbations offer
any
 solution to the first, basic cosmological problem, without the help
of
 inflation. On the other hand, each of these  mechanisms can be
successfully
 implemented in the context of inflationary cosmology \cite{MyBook}.
Moreover,
 inflationary cosmology offers many other possibilities which do not
exist in
 the standard hot Big Bang theory. The list of new possibilities
includes, in
 particular,   production of exponentially large domains with
slightly
 different energy density inside each of them, or with the same
density but
 with different number density of baryons, or with the same density
and
 composition but with different amplitudes of density perturbations,
etc.
 \cite{MyBook,KL}. Therefore, even if later it will be found that in
addition
 to flat adiabatic perturbations one needs strings or textures, or
 perturbations with a local maximum in the spectrum,  or even
something much
 more exotic, this by itself will not be a signal of  an
inconsistency of
 inflationary cosmology. On the contrary, it is much easier to find
new
 sources of density perturbations in the context of inflationary
cosmology
 than in the standard hot Big Bang theory.

Moreover, it seems that for a consistent realization of  the theory
of
perturbations produced by strings or textures, or by any other
mechanism, one
still needs inflation. Indeed, even if some as yet unspecified
quantum gravity
effects  at the Planck density will be able to solve the homogeneity,
isotropy,
horizon  and flatness problems   without any use of inflation (which
does not
seem likely), it is hardly possible that they will be able to solve
the
primordial monopole problem, since the monopoles in GUTs are produced
at a
temperature $T_c \sim  10^{15}$ GeV $\sim  10^{-4}\, M_p$, when
quantum gravity
effects are negligible. Therefore, until we  learn how to solve the
primordial
monopole problem without inflation, the  theory of strings and
textures
produced in high temperature phase transitions  in  non-inflationary
cosmology
will remain inconsistent.

The situation with strings and textures is not quite trivial even
within the
context of inflationary cosmology. Indeed, a typical critical
temperature of a
phase transition in cosmologically interesting theories of strings
and textures
is about $10^{16}$ GeV. It is extremely difficult (though not
impossible) to
reheat the universe up to such temperature after inflation
\cite{MyBook,Lyth,HB}, and it will require some additional fine
tuning of
parameters to make the reheating temperature smaller than the
critical
temperature of the phase transition producing monopoles.

Of course, one may pretend that the problem does not exist, or
suggest to
postpone its discussion because of ``our overwhelming ignorance''
\cite{NobTur}.  One may even claim  that  the theory of textures by
itself
``seems to match the explanatory triumphs of inflation''  \cite{ST}.
A  more
constructive approach is to face the problem and to use specific
possibilities
offered by inflation  to rescue strings and textures. Indeed,
inflationary
cosmology provides a simple mechanism which may lead to cosmological
phase
transitions during or after inflation, without any need of reheating
of the
universe. This mechanism is particularly natural in the chaotic
inflation
scenario  \cite{KL,Lyth,Chaot,Prim},  but it works in other versions
of
inflationary cosmology as well \cite{New},  and it can easily explain
why
strings, textures and some other useful topological or
non-topological defects
are produced, whereas monopoles are not.

To make our arguments more clear and, simultaneously, to discuss some
nontrivial examples of density perturbations in the standard Big Bang
theory
and in inflationary cosmology, we will consider here a simple
$O(N)$-symmetric
model of
an N-component field $\vec \Phi = \{\Phi_1, \Phi_2, ... , \Phi_N\}$,
$N > 1$,
with the effective potential
 \begin{equation}\label{1}
V(\Phi) = - {1\over 2}\, {M^2_{\Phi}} \Phi^2 +
{\lambda_\Phi\over 4}\, (\Phi^2 )^2 +  V_o  \ ,
\end{equation}
where $\Phi^2 = \Phi_1^2 + \Phi_2^2 +  ... + \Phi_N^2$; \ $V_o =
{M^4_{\Phi}\over 4\,\lambda_\Phi}$ is added to keep the vacuum energy
zero in
the absolute minimum of $V(\Phi)$.
At high temperature, the $O(N)$ symmetry of this theory is restored,
$\Phi =
0$. As the temperature decreases, a phase transition with  symmetry
breaking
takes place.
Soon after the phase transition,  the length of the isotopic vector
$\vec \Phi$
acquires the value $ v = M_\Phi/\sqrt{\lambda_\Phi}$. However, its
direction
may differ in  causally disconnected regions. Later on,  all vectors
tend to be
aligned inside each causally connected domain (i.e. inside each
particle
horizon), but they  cannot become aligned outside the particle
horizon.
Consequently, the field $\vec \Phi$ always remains inhomogeneous on
the horizon
scale
\begin{equation}\label{2}
R_H = 2\, H^{-1} =  \sqrt{3\, M_p^2 \over 2\, \pi\rho} \ .
\end{equation}
 For $N = 2$ this model describes  global strings \cite{ZelVil},
$N=3$ global
monopoles \cite{MyBook}, \cite{GlobVil}, $N = 4$ textures
\cite{Text}. For
larger $N$, there are no topological defects. However, for all $N >
1$ this
model produces  additional density perturbations with almost flat
spectrum
\cite{Vil,Dav,TurSper,Krauss} due to misalignment of the Goldstone
field on the
horizon scale. A somewhat better estimate of the scale of
inhomogeneity is just
$H^{-1}$, since it still takes  some  time of the order of $H^{-1}$
until the
field becomes homogeneous inside the horizon.  A typical  variation
of the
scalar field $\Phi$ on this scale can be estimated by $\sqrt {2}\ v$.
This
leads to an estimate of the  energy density in the gradients of the
fields
\begin{equation}\label{3}
\delta \rho \sim  {8 \, v^2  \pi\rho\over 3 \, M_p^2} \ .
\end{equation}
The relative amplitude of energy density of these fluctuations does
not depend
on the horizon scale,
\begin{equation}\label{4}
{\delta \rho \over \rho} \sim {8 \, \pi v^2 \over 3 \, M_p^2} \ .
\end{equation}
It gives the desirable value ${\delta \rho \over \rho} \sim 10^{-5}$
for $v
\sim  10^{16}$ GeV.
One should be a little bit more accurate though, since if gradients
of the
scalar field are the same everywhere, then the energy density is
strictly
homogeneous.
A more detailed study of this question performed in \cite{TurSper}
shows that
at large $N$ the amplitude of density perturbations is suppressed by
an
additional factor of $\sqrt{N}$, which slightly increases the
necessary value
of $v$.

We wish to note again, that we are discussing now essentially the
same
mechanism which is used in the theory of global strings, monopoles
and
textures. However, this mechanism is  more  general since it does not
require
existence of any topological defects. Moreover, one may expect that
in many
cases the contribution of the topological defects to density
perturbations will
be subdominant, since the probability of their formation in this
model is
suppressed by a large combinatorial factor.

Now let us study the phase transition in this  model in  more detail.
The phase
transition occurs
due to the high temperature corrections to the effective potential
(\ref{1})
\cite{[1]},
\begin{equation}\label{5}
\Delta V(\Phi) = {\lambda_\Phi T^2 \over 12} (N + 2) \ .
\end{equation}
This gives the critical temperature of the phase transition
\begin{equation}\label{6}
T_c = v \ \sqrt {12\over N + 2} \ .
\end{equation}
This quantity is of the same order as $v$, it does not depend on
$\lambda_\Phi$
and it only weakly depends on $N$. Thus, in order to have ${\delta
\rho \over
\rho} \sim 10^{-5}$ in this model, one should have the phase
transition at  an
extremely large temperature $T_c \sim v \sim 10^{16}$ GeV. This
temperature is
close to the grand unification scale,  but the critical temperature
in grand
unified models typically is one order of magnitude smaller,  $T_c
\sim 10^{15}$
GeV \cite{Grand}. This brings us back to the two problems mentioned
in the
beginning of the paper. In  non-inflationary  cosmology all our
achievements
will be brought down by the basic inconsistency of the cosmological
theory and
by the primordial monopole problem. In  inflationary cosmology it is
very hard
to reheat the universe up to the temperature $T > 10^{16}$ GeV
\cite{MyBook,Lyth,HB}, and  if we are able to do it, we will get all
our
monopoles back.

Still,  if one is prepared to pay a high price for a new type of
perturbations,
then in  inflationary cosmology this can at least  be achieved in an
internally
consistent way.
The most obvious possibility is to add to the model (\ref{1}) some
other fields
(but not gauge fields!), interacting with the field $\Phi$ with a
coupling
constant much larger than $\lambda_\Phi$. This will reduce the
critical
temperature in this model. Then one    tunes  the reheating
temperature to make
it smaller than the critical temperature in grand unified models but
larger
than the critical temperature in the extended model (\ref{1}).

There also exists  another, less trivial possibility, which has
certain
advantages \cite{KL,Lyth,Chaot,Prim}.
Let us assume that the inflaton field $\varphi$, which drives
inflation,
interacts with the field $\Phi$ with a small coupling constant $g^2$:
 \begin{equation}\label{7}
V(\varphi, \Phi) = {m^2_{\varphi} \over 4} \varphi^2 - {1\over 2}\,
{M^2_{\Phi}} \Phi^2 +
{\lambda_\Phi\over 4}\, (\Phi^2 )^2 + {g^2\over 2}  {\varphi^2}
\Phi^2  + V_o
\  .
\end{equation}
The inflaton mass should be sufficiently small, $m_\varphi \la
10^{13}$ GeV,
to make standard adiabatic perturbations produced during inflation
smaller than
$10^{-5}$ \cite{MyBook}. The effective mass of the field $\Phi$ at
$\Phi = 0$
depends on
$\varphi$:
 \begin{equation}\label{8}
{M^2_{\Phi}(\varphi)} =  - M^2_{\Phi} + g^2  {\varphi^2} \ .
\end{equation}
At large $\varphi$ the effective mass squared $M^2_{\Phi}(\varphi)$
is
positive and bigger than $H^2$. This means that at the beginning of
inflation,
when  the inflaton field $\varphi$ is extremely large, the $O(N)$
symmetry is
restored,  $\Phi = 0$. However, at $\varphi =\varphi_c$, where
\begin{equation}\label{9}
\varphi_c = {M_\Phi \over g}  = {\sqrt{2\lambda_\Phi} \, v \over g} \
,
\end{equation}
the phase transition with the
$0(N)$  symmetry breaking takes place. Thus, the  inflaton  field
$\varphi$
plays here the same role as the temperature in the standard theory of
phase
transitions. However, if it does not interact with  the Higgs fields,
which are
 responsible for the symmetry breaking in GUTs, its variation will
not lead to
any phase transitions   with  monopole production. Moreover, even if
monopoles
are produced, their density exponentially decreases after   the
phase
transition. Strings and textures will lead to important cosmological
effects
even if the universe inflates by a factor $e^{50}$ after the phase
transition,
whereas even much smaller inflation makes monopoles entirely
harmless.

A similar  mechanism may work in the new inflationary scenario as
well
\cite{New}.
However, in chaotic inflation  this mechanism is much more natural
and
efficient, since  the variation of the field $\varphi$  in this
theory is very
significant. At the last stages of inflation in our model, when the
structure
of the observable part of the universe was formed,  and after the
inflation,
when the inflaton field continued rolling down to the minimum of the
effective
potential, it  decreased by an extremely large value $\sim 3.5\times
10^{19}$
GeV,  from  $3.2\,  M_p$ to $0$ \cite{MyBook}. Correspondingly, the
effective
mass squared $M^2_{\Phi}(\varphi)$ changes by about $10 g^2 M_p^2$.

One cannot easily (without the help of supersymmetry) take the
constant $g^2$
in (\ref{7}) arbitrarily large, since radiative corrections would
induce an
extra term in the expression for   the effective potential
\cite{MyBook}:
\begin{eqnarray}\label{1a}
\delta V(\varphi) &\sim& N \  {M^4_{\Phi}(\varphi) \over 64\pi^2} \
\ln
{M^2_{\Phi}(\varphi)\over M^2_{\Phi}} \  \sim \ {N \,(g^2 \varphi^2 -
M^2_{\Phi})^2 \over 64\pi^2} \  \ln {g^2 \varphi^2 - M^2_{\Phi}\over
M^2_{\Phi}} \nonumber\\
  &\sim& {N \,g^4  \over 64\pi^2} \Bigl(\varphi^2 -
\varphi_c^2\Bigr)^2  \ln
{\varphi^2 - \varphi_c^2 \over \varphi_c^2}\ .
\end{eqnarray}
 In order to have ${\delta\rho\over \rho} \la10^{-5}$ for standard
inflationary
adiabatic perturbations generated in a theory with such an effective
potential,
 one should take $g^2 \la10^{-6}$.
For definiteness, let us take $g^2 \sim 10^{-7}$ and $m_\varphi \sim
10^{12}$
GeV. In this case we avoid large inflationary perturbations and make
the
additional term ${N \, g^4  \over 64\pi^2} \Bigl(\varphi^2 -
\varphi_c^2\Bigr)^2  \ln {\varphi^2 - \varphi_c^2 \over \varphi_c^2}$
smaller
than ${m_\varphi^2 \varphi^2\over  2}$ at the last stages of
inflation.
During the rolling of the field $\varphi$ from $\varphi \sim 3 \,
M_p$ to
$\varphi = 0$, the effective mass squared (\ref{8}) of the field
$\Phi$ changed
from  $- M^2_{\Phi} + \left(3\times 10^{16}\right)^2$ GeV$^2$ to $-
M^2_{\Phi}$. This  leads to the desired phase transition
for  $M_{\Phi} < 3\times 10^{16}$ GeV, which is certainly the case in
the
theory under consideration.

This scenario has a very interesting feature. The wavelength  of
perturbations,
which are produced  when the inflaton field is equal to $\varphi$,
later grows
up to $l(\varphi) \sim \exp(2\pi\varphi^2/M^2_p)$ cm   due to
inflation and
subsequent expansion of the universe   \cite{MyBook}. These
perturbations in
our model have a flat spectrum, but only on a scale $l \ga
l(\varphi_c) \sim
\exp(2\pi\varphi_c^2/M^2_p)$ cm. If the phase transition occurs at
$\varphi_c >
3.2 \,M_p$, all inhomogeneities produced by the gradients of the
field $\Phi$
will be stretched away from  the observable part of the universe.
Perturbations
produced at $2.8 \, M_p < \varphi_c < 3.2 \, M_p$ will form the
superlarge
structure of the observable part of the universe, but they will not
contribute
to perturbations on the galaxy scale. Finally, if the phase
transition occurs
at $\varphi_c \ll 2.8 \, M_p$, all observational consequences will be
the same
as if it were the ordinary finite temperature phase transition in the
theory
(\ref{1}).

For example, for $v \sim 10^{16}$ GeV,  $g^2 \sim
10^{-7}$,  $\lambda_\Phi = 0.5$, the phase transition occurs at
$\varphi_c \sim
3.2 \times 10^{19}$ GeV, and the corresponding density perturbations
only
appear on the horizon, at $l \ga \exp(2\pi\varphi_c^2/M^2_p) \sim
10^{28}$ cm.
For $\lambda_\Phi = 0.1$, perturbations with flat spectrum appear on
all
cosmologically interesting scales. By increasing $m_\varphi$ up to
about
$2\times 10^{12}$ GeV we obtain a mixture of the standard
inflationary
perturbations with ${\delta \rho \over \rho} \sim 10^{-5}$ and the
new ones.
The amplitude of each of these components is controlled by
$m_\varphi$ and $v$
respectively, and the cut-off of the spectrum of the new
perturbations on small
scales is controlled by  $\varphi_c = \sqrt{2 \lambda_\Phi}\, v/g$.
The model
describes both inflation and the phase transition in the theory
(\ref{1}), and
it  does not contain any coupling constants smaller than $10^{-7} -
10^{-6}$.
Since such coupling constants are known to exist even in the standard
model of
the electroweak interactions, this model seems to be sufficiently
natural.

Note, that at the end of inflation in this model, the field $\varphi$
still is
extremely large, $\varphi \sim M_p/5 \sim 2\times 10^{18}$ GeV.
Therefore, for
$\lambda_\Phi \la 10^{-2}$ the phase transition may
occur even after the end of inflation. This indicates that the
mechanism
discussed in this paper is rather general, and that  the field
triggering the
phase transition may differ from the field which drives inflation.
However,
chaotic inflation provides a particularly natural framework for the
realization
of this mechanism for generation of density perturbations.

Now let us try to see whether our  model admits any interesting
generalizations
and/or simplifications. An obvious idea is  to replace the field
$\Phi$ of the
model (\ref{1}) by the fields $\Phi$ and $H$ of the $SU(5)$ model
\cite{ShafVil,MyBook,Chaot}. There exist some reasons to do it. First
of all,
even though the models with broken {\it global} \, $O(N)$ symmetries
may have
interesting cosmological implications, so far there is no independent
reason to
 consider them  as a part of a realistic theory of elementary
particles.
Moreover, recently it was argued that quantum gravity corrections may
induce
large additional terms in the effective potential  (\ref{1}), which
will break
the $O(N)$ invariance \cite{XTextures}. If these terms lead to
existence of one
preferable direction in the isotopic space, they eliminate textures.
But if
they
lead to existence of several minima of equal depth, then domain walls
will be
produced after the phase transitions.  This  is a cosmological
disaster, which
can be avoided only if the universe inflates more than $e^{60}$ times
after the
phase transition.

Meanwhile, the model (\ref{7}), the fields $\Phi$ being interpreted
as the
$SU(5)$ Higgs fields in a {\it gauge} theory with a spontaneous
symmetry
breaking, represents the simplest semi-realistic  model of chaotic
inflation
\cite{MyBook,Chaot}. In such models we do not have textures, but we
may
have exponentially large strings. In addition to that, we may have
density
perturbations with a spectrum which grows on large scales, and then
becomes
flat on some scale $l > l_c$. Indeed, one can easily show that
standard
inflationary density perturbations generated in the model (\ref{7})
 on scale  $l \la \exp(2\pi\varphi_c^2/M^2_p)$ cm are smaller than
perturbations on  scale $l \ga \exp(2\pi\varphi_c^2/M^2_p) $ cm. The
reason is
the following. The amplitude of perturbations produced when the
inflaton field
was equal to $\varphi$ is given by \cite{MyBook}
\begin{equation}\label{10}
{\delta\rho\over\rho} = {48\over 5}\, \sqrt{2\pi\over 3} \
{V^{3/2}(\varphi)
\over M^3_p \ V^\prime (\varphi)} \ ,
\end{equation}
where $V^\prime (\varphi)$ is the derivative of the effective
potential, which
is responsible for the speed of rolling of the field $\varphi$.
Before the
phase transition $V^\prime (\varphi) = m^2_\varphi \varphi$. After
the phase
transition the field $\Phi$ rapidly falls down to the minimum of its
effective
potential at $v^2(\varphi) =  M^2_{\Phi}(\varphi)/\lambda_\Phi$.
Effective
potential {\it along this trajectory} is given by
\begin{equation}
V(\varphi) = {m^2_\varphi\over 2}  \varphi^2 + {1\over 4
\lambda_\Phi}\,
\Bigl[M_\Phi^4 - (M^2_\Phi - g^2 \varphi^2)^2\Bigr] \ .
\end{equation}
In the vicinity of the critical point $\varphi \sim \varphi_c =
M_\Phi/g$ the
modification of the effective potential by the last term is very
small, being
quadratic in $(M^2_\Phi - g^2 \varphi^2)$. However, the derivative of
the
effective potential at $\varphi < M_\Phi/g$ changes more
considerably,
\begin{equation}\label{2a}
V^\prime (\varphi) = m^2_\varphi \varphi + {g^2 \, \varphi \over
\lambda_\Phi}
\ (M^2_\Phi - g^2 \varphi^2) \ .
\end{equation}
The last term in (\ref{2a}) increases the speed of rolling of the
field
$\varphi$ and decreases the amplitude of density perturbations
generated after
the phase transition. The role of this term depends on the values of
parameters; in some cases it just decreases the amplitude of small
scale
density perturbations, in some other cases it may even lead to an
abrupt end of
inflation at the moment of the phase transition \cite{Axions}. Other
examples
of the spectra bending due to inflationary phase transitions can be
found in
\cite{KL,KP}.

Even if there are no phase transitions and topological defects in our
model
(e.g., if the sign of the term $M^2_\Phi \Phi^2$ in  (\ref{1}) is
positive),
inflation may still produce
 density perturbations with a non-flat spectrum   \cite{HBKP,HB}. To
give a
simple example, let us consider an effective potential, which, for
sufficiently
large $\varphi$, looks as follows:
\begin{equation}\label{11}
V(\varphi) = {m^2_\varphi \varphi^2\over 2} \ \left( 1 +
{\varphi^2\over
\varphi_o^2} \ln {\varphi\over \varphi_o} - {3\over 4}\
{\varphi^2\over
\varphi_o^2}\right) \ ,
\end{equation}
where $\varphi_o$ is some normalization parameter. Such  potentials
may appear
in the theory (\ref{7}) and in GUTs at $M_\Phi \ll g \varphi$ due to
radiative
corrections to $V(\varphi)$ \cite{MyBook}. This potential has a
minimum at
$\varphi = 0$, and it grows with an increase of the field $\varphi$
everywhere
except the point $\varphi = \varphi_o$, where $V^\prime = 0$. This
potential
may lead to small density perturbations produced during inflation at
$\varphi
\ll \varphi_o$ or at $\varphi \gg \varphi_o$, but, according to
(\ref{10}), it
has a very  high peak ${\delta\rho\over\rho} \ga 1$ corresponding to
fluctuations produced  near $\varphi = \varphi_o$. The height of the
peak can
be decreased by the decrease of the coefficient in front of the
logarithmic
term in (\ref{11}). The length scale corresponding to the maximum in
the
spectrum is controlled by the parameter $\varphi_o$. One should
emphasize, that
there is nothing special in this potential; even much more
complicated
potentials of this type are often discussed in the standard
electroweak theory
\cite{MyBook,sher}.

In the presence of the phase transition (of any type, not necessarily
leading
to textures), the effect discussed above is much   more natural and
pronounced.
Let us consider for example the effective potential which may appear
in the
model (\ref{7}) due to one loop radiative corrections (\ref{1a}):
\begin{equation}\label{1b}
V(\varphi) = {m^2_{\varphi} \over 2} \varphi^2  + N \ {g^4  \over
64\pi^2}
\Bigl(\varphi^2 - \varphi_c^2\Bigr)^2  \ln {\varphi^2 - \varphi_c^2
\over
\varphi_c^2}  + V_o  \  .
\end{equation}
The same potential may appear in GUTs, with $N$ being replaced by
some other
combinatorial coefficient. Note, that the second term has a maximum
at the
critical point $\varphi = \varphi_c$. This term may  lead to a large
modification of $V^\prime$. In the vicinity of the critical point, at
$\Delta\varphi = \varphi - \varphi_c \ll \varphi_c$, the effective
potential
(\ref{1b}) can be represented in the following form:
\begin{equation}\label{2b}
V(\varphi) = {m^2_{\varphi} \over 2} \varphi^2  + {N \,g^4
\varphi_c^2 \over
16\pi^2} (\Delta\varphi)^2   \ln {2\Delta\varphi\over \varphi_c}  +
V_o  \  .
\end{equation}
Let us take $m^2 \ll  {N \,g^4 \varphi_c^2 \over 8\pi^2}$. After some
elementary algebra, one can show that in this case the first
derivative of the
effective potential reaches its minimum at some point $\varphi_m =
\varphi_c +
\Delta\varphi$, where
${\Delta\varphi\over \varphi_c} \approx {1\over 2}\, e^{-3/2} \sim
0.1$. This
minimum will correspond to a maximum in the spectrum of
${\delta\rho\over\rho}$
at $\varphi = \varphi_m$. If one wishes, for example, to make this
maximum $C$
times higher than the value of ${\delta\rho\over\rho}$ at $\varphi =
\varphi_c$, one should take
\begin{equation}
m^2 \approx {N \,g^4 \varphi_c^2 \over 8\pi^2} \times {C \over 2(C -
1)}\,
e^{-3/2} \ .
\end{equation}
This is quite consistent with our assumption that $m^2 \ll  {N \,g^4
\varphi_c^2 \over 8\pi^2}$ for $C \ga 2$. For relatively small $C$,
the
width of the peak
is comparable with $\Delta\varphi  \sim 0.1 \, \varphi_c$.
For large $C$, the effective width of the peak becomes  smaller.
If the phase transition in our model occurs at
$\varphi_c  \sim 2.8 \, M_p$, then the maximum of the spectrum will
be
displaced at some point $\varphi_m$ in the interval $2.8 \, M_p <
\varphi < 3.2
\, M_p$. In other words, this spectrum grows $C$ times and then
decreases again
when the length scale changes from the galaxy scale to the scale of
horizon.
But this is exactly the type of the spectrum which  is necessary in
order to
improve the theory of formation of large scale structure of the
universe in the
cold dark matter model, and, simultaneously, to avoid an excessively
large
anisotropy of the microwave background radiation!

A detailed theory of this effect strongly depends on relations
between particle
masses and the Hubble parameter at the moment of the phase
transition. In some
cases, one may obtain a sharp maximum in the spectrum even without
any account
of the one loop corrections to the effective potential \cite{KP}.
However, the
fact that the one loop  contribution $\delta V$ typically has a
maximum near
the point of the phase transition (\ref{1a}), makes this effect more
general.

To summarize our results, inflationary phase transitions in GUTs
and/or in the
theories with a global symmetry breaking  may lead to production of
adiabatic
perturbations with a spectrum  which looks almost flat on very large
scale,
which has a relatively narrow maximum at $l(\varphi_m) \sim
\exp(2\pi\varphi_m^2/M^2_p)$ cm, and which decreases even further at
$l <
l(\varphi_c) \sim \exp(2\pi\varphi_c^2/M^2_p)$ cm.  In addition to
these
perturbations, on a scale $ l > l(\varphi_c)$ one may have the same
strings,
global monopoles, topological and nontopological textures which would
be
produced by  the ordinary high temperature phase transitions. The
amplitudes of
 inhomogeneities of all types and the values of length scales
$l(\varphi_m)$
and $l(\varphi_c)$ are controlled by values of  masses and coupling
constants
in the underlying theory of elementary particles.

One should remember also, that even the ordinary high temperature
phase
transition in the $SU(5)$ model occurs by a simultaneous production
of domains
of  four different phases: $SU(3)\times SU(2)\times U(1)$,
$SU(4)\times U(1)$,
$SU(3)\times (U(1))^2$ and $(SU(2))^2 \times (U(1))^2$ \cite{KShT}.
There is no
reason to expect that inflationary phase transitions are simpler. On
the
contrary, one may expect that the inflaton field couples differently
to
different scalar fields, which leads to a series of phase transitions
at some
critical values  $\phi_i$ of the inflaton field. This may create an
exponential
hierarchy of cosmological scales $l(\varphi_i) \sim
\exp(2\pi\varphi_i^2/M^2_p)$ cm.

These examples show  that inflation is extremely flexible and can
incorporate
all possible mechanisms of generation of density perturbations
\cite{MyBook}.
This does not mean that one can do whatever one wishes; for example,
it is very
hard to avoid the standard prediction that the density of the
universe should
be equal to critical. One should always keep in mind the possibility
that some
new observational data will contradict {\it all} versions of
inflationary
theory,
including the versions with strings, textures and non-flat spectra of
perturbations.  However, this did not happen yet. On the contrary,
one
may be afraid that  those who wish to have simple and definite
predictions to be compared with observations will feel embarrassed by
the
freedom of choice given to us by inflation. But do we ever have too
much
freedom? In order to understand
the present situation, let us try to draw some parallels with the
development
of the standard model of electroweak interactions.

The four-fermion theory of weak interactions had a very simple
structure, but
it was unrenormalizable. In the late 60's many scientists hoped that
one may
make sense out of this theory by performing a cut-off at the Planck
energy.
However, this theory had problems even on a much smaller energy scale
(violation of the unitarity bound on the electroweak scale).  just as
all
non-inflationary models had problems with monopoles on scales much
smaller than
$M_p$. The model suggested by Weinberg and Salam  \cite{WeinbSal} is
not
particularly simple; just remember how long it takes to write a
complete
Lagrangian. It has anomalies, which are to be cancelled.  It has
about
twenty adjustable parameters,  some of which look extremely
unnatural. For
example, most of the coupling constants are of order $10^{-1}$,
whereas the
coupling of the electron to the  Higgs boson is $G_e \sim 2\times
10^{-6}$. Therefore this model remained relatively unpopular for 4
years after
it was proposed, until it was realized that gauge theories with
spontaneous
symmetry breaking are renormalizable \cite{Hooft}.  Soon after that,
Georgi and
Glashow proposed an $O(3)$-symmetric theory of electroweak
interactions, which
was much simpler and which did not have any anomalies \cite{GG}.
Then, neutral
currents were discovered, which could not be described by this model.
However,
nobody considered the problems of the simplest models of electroweak
interactions as a signal of a general failure of gauge theories with
spontaneous symmetry breaking. The possibility to describe neutral
currents and
the existence of many adjustable parameters made the Weinberg-Salam
model
flexible enough to survive and to win. And now, instead of speaking
about
fine-tuning
of   parameters of this theory, we are speaking about measuring their
values.

This teaches us several interesting lessons. First of all, as
stressed by Abdus
Salam many years ago, Nature is not economical in particles, it is
economical
in principles. After we learned the principle of constructing {\it
consistent}
theories of elementary particles by using spontaneously broken gauge
theories,
there was no way back to the old models.  Similarly, after we learned
the
principle of constructing internally consistent cosmological models,
 it is very hard to forget it and return to
old cosmological theories.  In order to account for the abnormal
smallness of
density perturbations, ${\delta\rho\over \rho} \sim 10^{-5}$, one
should
consider theories with very small coupling constants. But the same
happened in
the electroweak theory, when, in order to account for the smallness
of the
electron mass, it was necessary to adjust the coupling constant
$G_e$ to be
$2\times 10^{-6}$. Nobody says now that this  is a fine tuning.
We expect that in the next decade observational cosmology  will
provide us with
a large amount of reliable data. It will be absolutely wonderful if
the
simplest version of inflationary cosmology with adiabatic
perturbations with a
flat spectrum is capable of describing all these data. One should
continue
investigation of this attractive possibility. However, there is no
special
reason to expect that the future cosmological theory will be much
simpler than
the  theory of electroweak interactions, with its twenty adjustable
parameters.
On the contrary, it may be extremely difficult to suggest any simple
theory
which would describe new observational data, see e.g. \cite{Gorski}.
One
should be prepared to most radical changes in the cosmological
theory,
but one should try to make these changes without breaking the
internal consistency of the theory. We believe
that the large flexibility of inflationary cosmology in providing
many
different sources of density perturbations is  a distinctive
advantage of this
theory. It is particularly interesting that most of the  sources of
nontrivial
density perturbations are related to inflationary phase transitions.
This
should make it possible to use cosmology as a powerful tool of
investigation of
 the phase structure of the elementary particle theory.

I am grateful to Rick Davis, Kris Gorski, Lev Kofman and David
Schramm   for
many useful discussions. This research was supported in part  by the
National
Science Foundation grant PHY-8612280.

\pagebreak

\end{document}